\documentclass[12pt]{iopart}
\usepackage[utf8]{inputenc}
\usepackage{amsfonts}
\usepackage{enumerate} 
\usepackage{graphicx}
\usepackage{xcolor}

\newcommand{\Alfven}{Alf\'ven }
\newcommand{\Poincare}{Poincar\'e }
\newcommand{\boldpsi}{\mathrm{\mathbf{\Psi}}}
\newcommand{\eqref}[1]{\eref{#1}}
\newcommand{\figref}[1]{Fig. \ref{#1}}
\newcommand{\Figref}[1]{Figure \ref{#1}}
\newcommand{\secref}[1]{Section \ref{#1}}
\newcommand{\appref}[1]{\ref{#1}}

\newcommand{\modi}{\color{black}}
\newcommand{\modii}{\color{black}}
\newcommand{\norm}{\color{black}}

\begin{document}

\bibliographystyle{Science}

\title{Stepped pressure equilibrium with relaxed flow and applications in reversed-field pinch plasmas}

\author{Z.S. Qu$^1$, R.L. Dewar$^1$, F. Ebrahimi$^2$, J.K. Anderson$^3$, S.R. Hudson$^2$, and M.J. Hole$^{1,4}$}
\address{$^1$Mathematical Sciences Institute, the Australian National University, Canberra ACT 2601, Australia}
\address{$^2$Princeton Plasma Physics Laboratory, PO Box 451, Princeton, New Jersey 08543, USA}
\address{$^3$Department of Physics, University of Wisconsin–Madison, Madison, Wisconsin 53706, USA}
\address{$^4$Australian Nuclear Science and Technology Organisation, Locked Bag 2001, Kirrawee DC NSW 2232, Australia}
\ead{zhisong.qu@anu.edu.au}
\vspace{10pt}

\begin{abstract}
The Multi-region Rela\underline{x}ed MHD (MRxMHD) has been successful in the construction of equilibria in three-dimensional (3D) configurations. 
In MRxMHD, the plasma is sliced into sub-volumes separated by ideal interfaces, each undergoing relaxation, allowing the formation of islands and chaos. 
The resulting equilibrium has a stepped pressure profile across sub-volumes.
The Stepped Pressure Equilibrium Code (SPEC) [S.R. Hudson \etal, Phys. Plasmas 19, 112502 (2012)]
was developed to calculate MRxMHD equilibria numerically.
In this work, we have extended the SPEC code to compute MRxMHD equilibria with field-aligned flow and rotation, 
following the theoretical development to incorporate cross-helicity and angular momentum constraints. 
The code has been verified for convergence and compared to a Grad-Shafranov solver in 2D.
We apply our new tool to study the flow profile change before and after the sawtooth crash of a reversed-field pinch discharge,
in which data of the parallel flow is available.
We find the promising result that under the constraints of cross-helicity and angular momentum, the parallel flow profile in post-crash SPEC equilibrium is flat in the plasma core and the amplitude of the flow matches experimental observations.
Finally, we provide an example equilibrium with a 3D helical field structure as the favoured lower energy state.
This will be the first 3D numerical equilibrium in which the flow effects are self-consistently calculated.
\end{abstract}

\section{Introduction}
%3D MHD
The solution of three dimensional magnetohydrodynamics (MHD) equilibria in toroidal confinement devices is a fundamental problem and an active area of research in fusion plasma physics.
In fact, the magnetic field line flow (following the magnetic field lines) is a $1\frac{1}{2}$ degree of freedom Hamiltonian dynamical system with $\nabla \cdot \mathbf{B}=0$,
where $\mathbf{B}$ is the magnetic field  \cite{Cary1983, Hudson2004}.
When the plasma equilibrium is axisymmetric, the toroidal angle $\varphi$ is an ignorable coordinate and therefore, according to Noether's theorem,
a constant of motion exist and the system is integrable.
In other words, all the field lines lie on nested axisymmetric, or two-dimensional (2D), toroidal surfaces known as the flux surfaces.
\modii Axisymmetry is broken in stellarators, in tokamaks with a resonate magnetic perturbation (RMP) \cite{Evans2004},
or in reversed-field pinches with 3D relaxed states (e.g.  \cite{Escande2000}). \norm
In general, for stellarator fields, only flux surfaces with a ``sufficiently'' irrational rotational transform exist (see for example,  \cite{Meiss1992,Lichtenberg1992} and references therein).
Magnetic islands form at surfaces where the rotational transform is rational.
As the deviation from axisymmetry increases, islands start to overlap, leading to regions with chaotic field lines.
The ideal MHD equilibrium $\nabla p = \mathbf{J} \times \mathbf{B}$, where $p$ is the pressure and $\mathbf{J}=\nabla \times \mathbf{B}$ the current density,
requires that $\mathbf{B} \cdot \nabla p = 0$, i.e. the pressure being a constant along the field line.
Therefore, the pressure profile is complicated with islands and chaos \cite{Grad1967}.

%Introduce MRxMHD.
%Introduce SPEC.
The Multi-region Rela\underline{x}ed MHD (MRxMHD) \cite{Bruno1996,Hole2007,Hole2009} 
considers a weak solution to the ideal MHD equilibrium by partitioning the plasma volume into a finite number of sub-domains separated by non-relaxed interfaces.
It sets the theoretical basis on a variational principle and is well-defined mathematically.
Within each volume, it seeks the minimum energy solution with the conservation of magnetic helicity $K$ given by
\begin{equation}
    K = \int \mathbf{A} \cdot \mathbf{B} dV,
\end{equation}
following the conjecture by Woltjer and Taylor \cite{Woltjer1960, Taylor1974, Taylor1986}, where $\mathbf{A}$ is the vector potential and $\mathbf{B} = \nabla \times \mathbf{A}$.
The resulting equilibrium has a constant pressure within each volume and the magnetic field is a linear Beltrami field $\nabla \times \mathbf{B} = \mu \mathbf{B}$ with $\mu$ the helicity multiplier.
Across the interfaces, the total force $p+B^2/2$ should be balanced.
The existence of solution for such a system is guaranteed by theorems of Bruno and Laurence  \cite{Bruno1996}.
Moreover, in axisymmetric systems MRxMHD can approach the continuously nested flux surfaces solution in the limit of $N_v \rightarrow \infty$, where $N_v$ is the number of interfaces \cite{Dennis2013a}.
It thus forms a bridge between the oversimplified Taylor relaxation and the infinitely constrained ideal MHD with nested flux surfaces.
To access to MRxMHD equilibrium solutions with complicated geometry and parameters, the Stepped-Pressure Equilibrium Code (SPEC) \cite{Hudson2012a,Hudson2012b} was built and verified in stellarator geometry \cite{Loizu2016} .
It has been applied to resolve current sheets \cite{Loizu2015, Loizu2015a}, tearing modes \cite{Loizu2019}, RMP \cite{Loizu2016b}, and beta-limits in stellarators \cite{Loizu2017}.
A time-dependent version of MRxMHD is being developed to study waves and instabilities of the equilibrium state \cite{Dewar2015,Dewar2017,Dewar2019}.

%Introduce why we add flow and Graham's paper.
\modii For tokamaks, the neutral beam injection introduces external torque, which drives flows  \cite{Suckewer1979} that could co-exist with error field from RMP. \norm
Although flow is neoclassically damped in general stellarators, the introduction of quasi-symmetry could allow undamped flow to exist \cite{Spong2005,Helander2008,Simakov2011}.
This has been confirmed experimentally on the HSX stellarator \cite{Kumar2018,Dobbins2019}.
Such equilibria are intrinsically 3D and with flow.
\modii However, the effect of flows has not be present in 3D equilibrium codes \norm
(except for HINT \cite{Suzuki2006} and SIESTA \cite{Hirshman2011}, which can be better described as initial value codes solving the resistive MHD evolution equations). 
\modii In this work, we present an implementation of 3D equilibrium solver based on MRxMHD with flow. \norm
To introduce flow in the frame work of variational principle,
\modii we seek \norm a state with minimum energy (or equivalently maximum entropy) subject to a global constraint known as the cross-helicity defined by
\begin{equation}
    C = \int \mathbf{u} \cdot \mathbf{B} dV,
\end{equation}
where $\mathbf{u}$ is the fluid velocity.
The resulting state has $\mathbf{u}$ parallel to $\mathbf{B}$ and thus was given the name ``dynamic alignment'' \cite{Matthaeus1983} by the space plasma community to contrast the ``selective decay'' conjectured by Taylor \cite{Taylor1974},
which has energy decaying much faster than magnetic helicity.
Numerical experiments \modii later \norm confirmed the existence of such states as end states of turbulent space plasmas \cite{Stribling1991,Matthaeus2008}.
In toroidal confinement, Finn and Antonsen \cite{Finn1983} proved the conservation of cross-helicity for a barotropic, dissipation-free plasma and found the corresponding minimum energy state with both $C$ and $K$ globally conserved.
They also added the global constraint of toroidal angular momentum for axisymmetric machines which leads to toroidal rotations.
\modii An alternative approach to reach to similar \norm state is by accessing the limit of zero ion-skin-depth \cite{Hameiri2013,Hameiri2014} from the more general two-fluid Hall-MHD relaxation or double-Beltrami solution \cite{Mahajan1998,Yoshida2002,Lingam2016}.
The third approach is to consider the state as the stationary solution of time-dependent MRxMHD  \cite{Dewar2019}.

Dennis \etal \cite{Dennis2014} generalized Finn and Antonsen \cite{Finn1983} to include multiple volumes, forming the theoretical basis of MRxMHD with flow which will be used in this work.
In addition to the stepped pressure, it also predicts a stepped flow in toroidal and parallel direction,
with $\mathbf{u} = \lambda \mathbf{B} / \rho + \Omega R \hat{\mathbf{e}}_\varphi$,
where $\rho$ is the plasma mass density, $R$ the major radius, $\hat{\mathbf{e}}_\varphi$ the unit vector in the toroidal direction, and $\Omega$ and $\lambda$ are constants.
The form of the flow formulated by Dennis \etal is not general enough in many applications.
For example, it does not capture the $\mathbf{E} \times \mathbf{B}$ flow.
Moreover,  flow is discontinuous on interfaces without any drag force due to the absence of viscosity, leading to the well known D'Alembert's paradox \cite{Landau1987}.
Despite the limitations, we argue that it could be a step towards mathematically well-defined equilibria with a more general form of flow, and therefore there is value to build a 3D equilibrium code based on it.
Also, this equilibrium solution is essential for the further development of a time-dependent MRxMHD code.
The focus of this paper will be to \modii develop \norm such an equilibrium code,
\modii and apply it to relevant equilibria. 
Besides, we aim to demonstrate the physical implications of the MRxMHD model to equilibrium flow. 
\norm

%Introduce RFP and motivation.
\modii
The reversed-field pinch (RFP) has axisymmetric field coils similar to a tokamak and its poloidal field is generated mainly by plasma current.
Unlike tokamak, the toroidal field of RFP is weak, decreases as a function of minor radius and changes its sign at the plasma edge.
The field reversal is due to the measured dynamo effect \cite{Ennis2010} from tearing mode activities.
RFP in general has exhibited a richer class of 3D magnetic field configurations due to dynamo effects and self-organisation (e.g.  \cite{Escande2000}).
A major objective is to find whether the observed experimental relaxed states can be understood and predicted via theoretical relaxation models.
As a successful example, the equilibrium field bifurcation between the single-helical-axis state and the double-axis state in the RFX experiment was successfully reproduced by the static MRxMHD model and SPEC, considering partial relaxation \cite{Dennis2013}.
On the other hand, relaxation of axisymmetric flows and magnetic fields in RFPs have been theoretically shown to be interlinked \cite{Hegna1996}. 
In particular, during tearing mode reconnection events \cite{Choi2006} momentum transport and flow relaxation have been demonstrated using nonlinear MHD calculations of tearing mode torques  \cite{Ebrahimi2007,Ebrahimi2008} and later in two-fluid model \cite{sauppe2017}.
Experimentally, the parallel flow profile on Madison Symmetric Torus (MST), a RFP device, 
is flat from centre to edge then flips its sign at the edge during a reconnection event \cite{Kuritsyn2009}.
A relaxation model with flow may be able to reach such a relaxed flow profile.
Khalzov\etal \cite{Khalzov2012} applied the relaxed single-fluid and Hall-MHD models in a single-volume cylinder to study the case theoretically, and found that single-fluid MHD relaxation theory to be in reasonable agreement with experimental observation. However, their study was limited to  a single-volume cylinder with only cross-helicity conservation.
%Using MRxMHD with flow, Dennis \etal modeled the same case analytically with two volumes, also in cylindrical geometry. 
%The change of sign in the parallel flow profile was attributed to incomplete relaxation \textcolor{magenta}{by Khalzov \etal but was not pursued there due to the limitation of the single-volume model}.
In this work, we follow this idea and present a numerical calculation of the post-sawtooth-crash flow profile from pre-crash profile using our toroidal SPEC-flow code
which recovers the experimentally observed parallel flow profile both qualitatively and quantitatively.
We also construct the first multi-volume solution of the aforementioned MST case with 3D magnetic fields.
This will form the application part of the paper.
\norm

The work is organized as follows. 
\secref{sec:theory} reviews and clarified the theory of MRxMHD with flow.
\secref{sec:SPEC-Flow} explains the numerical details of SPEC with flow and provides a convergence study in stellarators and a benchmark with a tokamak equilibrium code.
\modi
\secref{sec:relax} computes the flow profile after sawtooth crash from a pre-crash equilibrium using SPEC and relaxation theory.
\secref{sec:3D} gives a numerical example of the stepped flow case with 3D magnetic fields in MST.
\norm
Finally, \secref{sec:conclusion} draws the conclusion.

\section{Theory} \label{sec:theory}
\subsection{Basic equations}
\begin{figure}[!htbp]
    \centering
    \includegraphics[width=8.5cm]{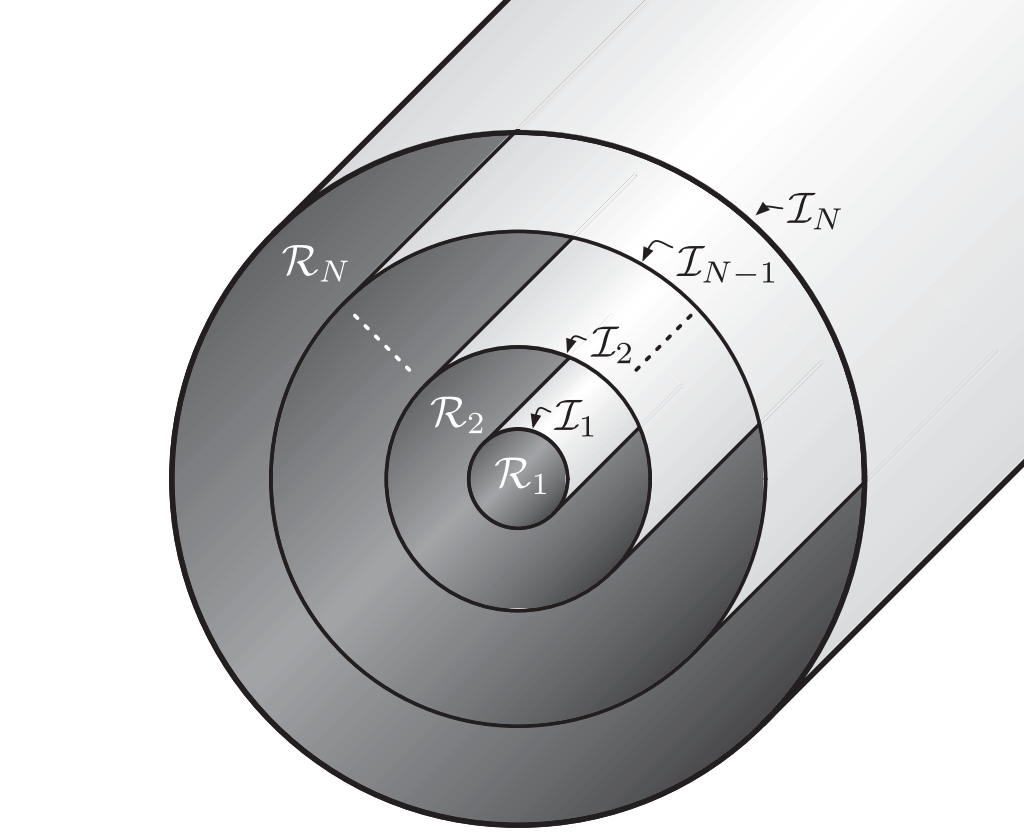}
    \caption{Schematic view of the plasma regions $\mathcal{R}_i$ and interfaces $\mathcal{I}_i$.}
    \label{fig:geometry}
\end{figure}

In this work, we mainly follow theoretical work of Dennis \etal \cite{Dennis2014}.
The plasma volume is separated into $N_V$ volumes labeled by $\mathcal{R}_i$. Between two volumes, there is an ideal interface labelled by $\mathcal{I}_i$. A schematic view of the magnetic geometry is shown in \figref{fig:geometry}.

Within each volume, the plasma is assumed to relax to the Taylor state, i.e. the plasma energy is minimized, subject to a number of constraints.
The total energy $E_i$ in each volume is given by
\begin{equation}
    E_i = \int_{\mathcal{R}_i} \left[ \frac{1}{2} \rho \mathbf{u}^2  + \frac{1}{2} \mathbf{B}^2 + \frac{p}{\gamma - 1} \right] d V,
\end{equation}
where $\gamma$ is the adiabatic index and the unit is CGS.
The integration is over the plasma volume $\mathcal{R}_{i}$.

There are a number of constrained quantities during the energy minimization.
They are the magnetic helicity $K_i$, the cross-helicity $C_i$,
the angular momentum $L_i$, the total mass $M_i$ and the plasma entropy $S_i$. 
These constraints are given by
\begin{eqnarray}
    \label{eq:magnetic_helicity}
    K_i =& \frac{1}{2} \int_{\mathcal{R}_i} \mathbf{A} \cdot \mathbf{B} dV, \\
    \label{eq:cross_helicity}
    C_i =& \int_{\mathcal{R}_i} \mathbf{u} \cdot \mathbf{B} dV, \\
    \label{eq:angular_momentum}
    L_i =& \int_{\mathcal{R}_i} \rho R^2 \mathbf{u} \cdot \nabla \varphi dV, \\
    \label{eq:total_mass}
    M_i =& \int_{\mathcal{R}_i} \rho dV, \\
    \label{eq:total:entropy}
    S_i =& \int_{\mathcal{R}_i} \frac{\rho}{\gamma -1} \ln \kappa \frac{p}{\rho^\gamma} dV, 
\end{eqnarray}
respectively, with $\kappa$ a constant.
In fact, $K_i$ is not fully gauge invariant, i.e. $K_i$ is not a constant under the gauge transformation $\mathbf{A} \rightarrow \mathbf{A} + \nabla g$ for multivalued $g$. 
However, in this work we will choose a specific gauge described in the numerical section and limit $g$ to be single-valued.
The angular momentum constraint is only applied to toroidal geometry with coordinates $(R, \varphi, Z)$, where $\varphi$ is the toroidal angle.
We note that unlike Dennis \etal, we add separate constraints for total mass and entropy.

The total energy functional of the entire plasma is given by
\begin{equation}
    W = \sum_{i=1}^{N_V} W_i.
\end{equation}
Within each volume we have
\begin{eqnarray}
   \label{eq:energy_functional}
\fl    W_i = E_i - \mu_i(K_i-K_{0i}) - \lambda_i(C_i - C_{0i})  \nonumber\\
    - \Omega_i(L_i-L_{0i}) - \nu_i(M_i - M_{0i}) - \tau_i (S_i - S_{0i}),
\end{eqnarray}
The quantities $\mu_i$, $\lambda_i$, $\Omega_i$, $\nu_i$ and $\tau_i$ are Lagrange multipliers,
while $K_{0i}, C_{0i}, L_{0i}, M_{0i}$ and $S_{0i}$ are constraint values of $K_{i}, C_{i}, L_{i}, M_{i}$ and $S_{i}$, respectively.
In addition, the poloidal and toroidal magnetic fluxes, $\psi_{p,i}$ and $\psi_{t,i}$, are taken as constraints.

The equilibrium of MRxMHD locates at the stationary points of the energy functional $W$,
and can be obtained from the variational principle.
The free variables are $p$, $\rho$, $\mathbf{u}$, $\mathbf{A}$, the Lagrange multipliers and the position of the interfaces.
Varying $W$ with respect to $p$ and setting the varied equation to zero, one gets the equation of state
\begin{equation}
    p = \tau_i \rho.
    \label{eq:eq_of_state}
\end{equation}
Comparison with the ideal gas law $p = n k_B T$ shows the physical meaning of $\tau_i$ to be $k_B T_i/m_i$,
where $n$ is the number density, $k_B$ the Boltzmann constant, $T_i$ the plasma temperature in the $i$'th volume.
Varying with respect to $\rho$, we reach the Bernoulli equation given by
\begin{equation}
    \tau_i \ln \frac{\rho}{\rho_{0i}} + \frac{1}{2} \mathbf{u}^2 - \Omega_i R^2 \mathbf{u} \cdot \nabla \varphi  = 0,
    \label{eq:bernoulli_original}
\end{equation}
where
\begin{equation}
    \ln \rho_{0i} = -\ln \kappa \tau_i + \frac{\gamma}{\gamma-1} - \frac{\nu_i}{\tau_i}.
\end{equation}
Varying with respect to $\mathbf{u}$, we get flow in the form of
\begin{equation}
    \rho \mathbf{u} = \lambda_i \mathbf{B} + \rho \Omega_i R^2 \nabla{\varphi}.
    \label{eq:u}
\end{equation}
Varying $\mathbf{A}$ leads to the modified Beltrami equation given by
\begin{equation}
    \nabla \times \mathbf{B} = \mu_i \mathbf{B} + \lambda_i \nabla \times \mathbf{u}.
    \label{eq:premitive_Beltrami}
\end{equation}
Finally, varying with respect to the interfaces, one obtains the interface jump condition
\begin{equation}
    \left[\left[ p + \frac{1}{2} B^2\right]\right] =0,
\end{equation}
where $[[x]]=x^- - x^+$ stands for the difference of $x$ on either side of the interface.

In addition, $\mathcal{I}_i$ are ideal interfaces with infinite conductivity, requiring that on these interfaces
\begin{equation}
    \mathbf{B} \cdot \mathbf{n} = 0,
    \label{eq:b_normal}
\end{equation}
where $\mathbf{n}$ is the unit vector perpendicular to the interfaces.
Also, any electrostatic potential difference on the interface would be short-circuited.
The equilibrium electric field (if there is any) should be perpendicular to the interfaces.
From the ideal Ohm's law operating on the interfaces, we have
\begin{equation}
    \mathbf{u} \cdot \mathbf{n} = 0, 
    \label{eq:u_normal}
\end{equation}
i.e. the plasma must not flow out of the volume.

Combining \eqref{eq:u}, \eqref{eq:b_normal} and \eqref{eq:u_normal},
one will reach that on the interfaces $\Omega_i \nabla \varphi \cdot \mathbf{n}  = 0$.
For a interface set by $F(R, Z, \varphi)=0$, the normal direction is parallel to $\nabla F$.
The aforementioned condition requires either $F = F(R, Z)$ (the interface is axisymmetric), or $\Omega_i =0$ (there is no rigid toroidal rotation).
As a conclusion, $\Omega_i$ can only be non-zero for volumes bounded entirely by axisymmetric interfaces.
This is consistent with the simulation finding on MAST that the introduction of RMP field will suppress toroidal flow  \cite{Liu2012}.
We note that it provides a more strict criterion than Dennis \etal,
where only the outermost boundary is required to stay axisymmetric.
Dennis \etal also proposed a scenario in which the interfaces position are time-dependent and rotating altogether with the plasma, but becoming time-independent in the rotating frame. 
On contrary, we consider all interfaces in our work to be static and will not consider the time-dependent case.

After some algebra, \eqref{eq:bernoulli_original}-\eqref{eq:u_normal} are summarized as follows.
\begin{eqnarray}
    \label{eq:Beltrami}
    \mathcal{R}_i&: \ \  \nabla \times \left( 1 - \frac{\lambda_i^2}{\rho} \right) \mathbf{B} = \mu_i \mathbf{B} + 2 \lambda_i \Omega_i \nabla{Z}, \ \ \mathrm{(Beltrami)} \\
    \label{eq:Bernoulli}
    \mathcal{R}_i&: \ \ \tau_i \ln \frac{\rho}{\rho_{0i}} + \frac{1}{2} \frac{\lambda_i^2 B^2}{\rho^2} = \frac{1}{2} \Omega_i^2 R^2, \ \ \mathrm{(Bernoulli)} \\
    \label{eq:Pressure_jump}
    \mathcal{I}_i&: \ \  \left[\left[ \rho \tau_i + \frac{1}{2} B^2\right]\right] =0, \ \ \mathrm{(Force\ balance)}\\
    \label{eq:Ideal_interface}
    \mathcal{I}_i&: \ \ \mathbf{B} \cdot \mathbf{n} = 0,  \ \ \mathrm{(Ideal\ interface\ condition)}
\end{eqnarray}
while $p$ and $\mathbf{u}$ are derived from \eqref{eq:eq_of_state} and \eqref{eq:u} once \eqref{eq:Beltrami}-\eqref{eq:Ideal_interface} are solved.
In additional, $\Omega_i \neq 0$ is allowed if the geometry is toroidal,
and if both $\mathcal{I}_{i-1}$ and $\mathcal{I}_i$ are axisymmetric.

\subsection{Continuous nested flux surfaces limit and comparison to tokamak equilibrium}
\label{sec:nested_flux_surfaces}
In this section, we will revisit the continuous nested flux surfaces limit of MRxMHD, using slightly different argument from Dennis \etal.
Straightforwardly, in the limit of continuous nested flux surfaces,
\eqref{eq:eq_of_state}, \eqref{eq:u} and \eqref{eq:Bernoulli} become
\begin{equation}
    p = \tau(s) \rho,
    \label{eq:eq_of_state_fs}
\end{equation}
\begin{equation}
    \mathbf{u} = \frac{\lambda(s)}{\rho} \mathbf{B} + \Omega(s) R^2 \nabla \varphi,
    \label{eq:u_fs}
\end{equation}
and
\begin{equation}
    \tau(s) \ln \frac{\rho}{\rho_0(s)} + \frac{1}{2} \left[\frac{\lambda(s) B}{\rho} \right]^2 = \frac{1}{2} [\Omega(s) R]^2,
    \label{eq:Bernoulli_fs}
\end{equation}
respectively, with ``s'' the continuous flux surfaces label (for example the square root of the normalized poloidal or toroidal flux).
One can compare these results with that of the ideal MHD in tokamaks geometry (e.g. Refs \cite{Iacono1990, McClements2010}).
The Grad-Shafranov-Bernoulli system of equations is summarized in \appref{sec:GSE}.
The forms of flow \eqref{eq:u_fs} and the Bernoulli equation \eqref{eq:Bernoulli_fs} are identical to that of the ideal MHD given by \eqref{eq:appendix_u} and \eqref{eq:appendix_Bernoulli}, respectively.

For continuous nested surfaces, \eqref{eq:u} and \eqref{eq:premitive_Beltrami} approach 
\begin{eqnarray}
\fl    \rho \mathbf{u} \cdot \nabla \mathbf{u} = - \nabla p + \mathbf{J} \times \mathbf{B} 
     + \rho \Omega(s) \nabla (R^2 \mathbf{u} \cdot \nabla \varphi) \nonumber\\
   - \rho \Omega(s) R^2 \nabla \varphi \times (\nabla \times \mathbf{u}),
   \label{eq:mhd_fs}
\end{eqnarray}
where $\mathbf{J} = \nabla \times \mathbf{B}$ is the current density.
The rigorous derivation is provided in Section III of Dennis \etal \cite{Dennis2014}.
Expanding the last two terms of \eqref{eq:mhd_fs} in cylindrical geometry,
one will find that for $\Omega \ne 0$, \eqref{eq:mhd_fs} can match that of the ideal MHD 
if and only the last two terms cancel.
This means the flow is axisymmetric.
In fact, this criteria is always satisfied by recalling that $\Omega_i \ne 0$ is only allowed if the bounding interfaces are axisymmetric.
Consequently,  the last two terms of \eqref{eq:mhd_fs} will always vanish and \eqref{eq:mhd_fs} conforms that of the ideal MHD force balance given by
\begin{equation}
    \rho \mathbf{u} \cdot \nabla \mathbf{u} = - \nabla p + \mathbf{J} \times \mathbf{B}.
    \label{eq:equation_of_motion}
\end{equation}

\section{Stepped pressure equilibrium code with flow} \label{sec:SPEC-Flow}
\subsection{Numerical methods}

\subsubsection{The static SPEC code}

The current version of SPEC solves the fixed-boundary MRxMHD equilibrium in slab, cylindrical or toroidal geometry.
Setting $\lambda_i = \Omega_i = 0$, \eqref{eq:Beltrami}-\eqref{eq:Ideal_interface} will reduce to the set of equations coded into SPEC, with \eqref{eq:Bernoulli} degenerated to $\rho = \rho_{0i}$ and replaced by the piecewise pressure $p_i = \rho_{0i} \tau_i$.
For toroidal geometry with stellarator symmetry, the interfaces and plasma boundary are specified by
\begin{eqnarray}
    \label{eq:parametric_R}
    R_{\mathcal{I}_{i}} &= \sum_{m,n} R_{m,n,i} \cos(m \theta - n N_p \zeta), \\
    \label{eq:parametric_Z}
    Z_{\mathcal{I}_{i}} &= \sum_{m,n} Z_{m,n,i} \sin(m \theta - n N_p \zeta), \\
    \varphi &= \zeta,
\end{eqnarray}
where $\theta$ and $\zeta$ are generalized angles.
Here, $N_p$ is the field periodicity.
The number of the poloidal and toroidal harmonics are known as $M_{\mathrm{pol}}$ and $N_{\mathrm{tor}}$, respectively.
The summation is over $n \in [-N_{\mathrm{tor}}, N_{\mathrm{tor}}]$ and $m \in [0, M_{\mathrm{pol}}]$
for $m,n \in \mathbb {N}$.
For the plasma boundary, $R_{m,n,N_V}$ and $Z_{m,n,N_V}$ are given as input parameters,
while they are unknowns for the interfaces and will be determined by the force balance condition \eqref{eq:Pressure_jump}.
Inside each volume, the coordinates are specified as a linear interpolation between the inner and outer interfaces, given by
\begin{eqnarray}
    R(s, \theta, \zeta) &= \frac{1-s}{2} R_{\mathcal{I}_{i-1}}(\theta_, \zeta) + \frac{1+s}{2} R_{\mathcal{I}_{i}}(\theta_, \zeta) \\
    Z(s, \theta, \zeta) &= \frac{1-s}{2} Z_{\mathcal{I}_{i-1}}(\theta_, \zeta) + \frac{1+s}{2} Z_{\mathcal{I}_{i}}(\theta_, \zeta),
\end{eqnarray}
with $s\in[-1,1]$ the generalized radial coordinate.
A special treatment is needed for the innermost volume and is described in Hudson \etal \cite{Hudson2012a}.

The vector potential $\mathbf{A}$ in SPEC takes the Clebsch form 
$\mathbf{A} = A_\theta \nabla \theta + A_\zeta \nabla \zeta$, with the two components represented by
\begin{eqnarray}
    A_\theta &= \sum_{m,n,l} A_{\theta,m,n,l} T_{l}(s) \cos(m \theta - n N_p \zeta), \\
    A_\zeta &= \sum_{m,n,l} A_{\zeta,m,n,l} T_{l}(s) \cos(m \theta - n N_p \zeta),
\end{eqnarray}
where $T_l(s)$ is the Chebyshev polynomial of order $l$, with $L_{\mathrm{rad}}$ the highest order number.
SPEC uses a gauge such that $\mathbf{A} = 0$ on the inner surface for volumes $\mathcal{R}_{i\ge2}$ and on the outer surface for $\mathcal{R}_1$.
The boundary condition $\mathbf{B} \cdot \mathbf{n}=0$, the enclosed poloidal flux $\Delta \psi_p$ and the toroidal flux $\Delta\psi_t$ are enforced by introducing another set of Lagrange multipliers ($e_i$ for the $i$'th Fourier harmonic of the boundary condition, $g,h$ for flux).
Let $\mathbf{a} = (A_{\theta,m,n,l}, A_{\zeta,m,n,l}, e_i, g, h)$,
the energy functional can be written as a quadratic form in $\mathbf{a}$, simply that
\begin{equation}
    W_{\mathrm{static}} = \frac{1}{2} \mathbf{a}^{\mathrm{T}} \cdot (\mathcal{A} - \mu_i \mathcal{D}) \cdot \mathbf{a} 
    - \mathbf{a}^{\mathrm{T}} \cdot \mathcal{B} \cdot \boldpsi + \mu_i K_{0i},
    \label{eq:static_quadratic_form}
\end{equation}
where $\mathcal{A}$, $\mathcal{B}$ and $\mathcal{D}$ are matrices constructed from the prescribed geometry,
and $\boldpsi = (\Delta \psi_p, \Delta\psi_t)$.
The solution $\mathbf{a}$ is the stationary point of \eqref{eq:static_quadratic_form}.
If $\mu_i$ is known and is not one of the eigenvalues of the eigenvalue problem 
$\mathcal{A} \cdot \mathbf{a} = \mu_i \mathcal{D}\cdot \mathbf{a}$,
the bracket term in \eqref{eq:static_quadratic_form} is invertible.
The solution is then given by $\mathbf{a} = (\mathcal{A} - \mu_i \mathcal{D})^{-1} \mathcal{B} \cdot \boldpsi$.
If $\mu_i$ is unknown but $K_{0i}$ is known, or $\mu_i$ is one of the eigenvalues,
the solution $\mathbf{a}$ should be obtained by finding the stationary point of \eqref{eq:static_quadratic_form} with
both $\mu_i$ and $\mathbf{a}$ as variables.

After the Beltrami field is solved in each volume,
SPEC will move the position of the interfaces according to force difference on the two sides using a Newton's method.
Finally, the force balance condition \eqref{eq:Pressure_jump} will be satisfied down to machine precision.

\subsubsection{Adding flow to SPEC}
Equation \eqref{eq:Beltrami}-\eqref{eq:Ideal_interface} are the system of equation to be solved in SPEC with flow. 
For a MRxMHD problem, one needs to adjust the Lagrange multipliers to satisfy the constraints in each volume, and the position of the interfaces to satisfy the force balance.
However, as SPEC is an equilibrium code, we also provide the option to treat the constants $\rho_{0i}$, $\kappa_i$, $\lambda_i$ and $\Omega_i$ as user inputs, 
since these quantities have more physical and measurable meanings ($\rho_{0i}$ is the density constant, $\kappa_i$ is $k_B T_i/m_i$, $\lambda_i$ related to the \Alfven Mach number, $\Omega_i$ the angular rigid rotation frequency).

One can substitute \eqref{eq:eq_of_state} and \eqref{eq:u} into the energy functional \eqref{eq:energy_functional}, giving that
\begin{eqnarray}
\label{eq:energy_functional_substitute}
\fl    W_i = \int_{\mathcal{R}_i} \left[\frac{1}{2}\left(1-\frac{\lambda_i^2}{\rho}\right) \mathbf{B}^2
    -\lambda_i \Omega_i R^2 \mathbf{B} \cdot \nabla \varphi \right] d V \nonumber\\
    - \mu_i (K - K_{0i}) + W_{\rho i}(\rho, \Omega_i, \kappa_i, \tau_i),
\end{eqnarray}
where $W_{\rho i}$ depends on $\rho$ and the Lagrange multipliers.
Taking $\rho$ as known, solved from \eqref{eq:Bernoulli} and discarding $W_{\rho i}$,
$W_i$ is transformed in to a matrix form given by
\begin{equation}
    W_{\mathrm{flow}} = W_{\mathrm{static}} - \lambda_i \Omega_i \mathbf{g}^{\mathrm{T}} \cdot \mathbf{a}, 
    \label{eq:flow_quadratic_form}
\end{equation}
where the second term corresponds to the second term in the bracket in \eqref{eq:energy_functional_substitute}
and the factor in front of $\mathbf{B}^2$ is now absorbed into $\mathcal{A}$,
making $\mathcal{A} = \mathcal{A}(\rho)$.
We can thus solve $\mathbf{a}$ using either a linear algebra method,
giving $\mathbf{a} = (\mathcal{A} - \mu_i \mathcal{D})^{-1} (\mathcal{B} \cdot \boldpsi + \mathbf{g}$),
or Newton's method by locating the stationary point of \eqref{eq:flow_quadratic_form}.

The next step is to obtain $\rho$ from the Bernoulli equation \eqref{eq:Bernoulli},
which is rewritten in the following form
\begin{equation}
    f(M_\parallel^2) = -\ln \frac{M_\parallel^2}{M_{\parallel 0}^2} + M_\parallel^2 - \frac{\Omega_i^2 R^2}{\tau_i} = 0,
    \label{eq:fm2}
\end{equation}
where $M_\parallel = \lambda_i B / \rho \sqrt{\tau_i}$ is the parallel Mach number, and $M_{\parallel 0} = \lambda_i B / \rho_0 \sqrt{\tau_i}$. 

The function $f$ has its minimum at $M_\parallel^2=1$. In general, if $f(1) < 0$,the equation $f(M_\parallel^2)=0$ can have two solutions: a subsonic solution $(0<M_\parallel^2<1)$ and a supersonic solution $(M_\parallel^2>1)$. 
This bifurcation in density was discussed in detail by Finn and Antonsen \cite{Finn1983}. It is possible to develop a density discontinuity (shock) \cite{Betti2000} in the system when $f(1)=0$ appears within the volume, i.e. the Bernoulli equation has two degenerated solutions $M_\parallel^2=1$. However, in this work, we have limited the discussion to purely subsonic or purely supersonic flow within each volume. We will not allow a transonic surface to develop in the volume (but it can happen across the interfaces).

The equations in each volume are solved in an iterative manner. 
Concretely, the solver takes the following steps:
\begin{enumerate}[(i)]
    \item Solves the (modified) Beltrami equation \eqref{eq:Beltrami} assuming $\rho = \rho_{0i}$.
    \item Solves the Bernoulli equation \eqref{eq:Bernoulli}, given $\mathbf{B}$ from the Beltrami solver.
    \item Solves \eqref{eq:Beltrami}, given $\rho$ from (ii).
    \item Repeat (ii) (iii) until converge.
\end{enumerate}
The iterative procedure converges to machine precision usually within five steps.
After the field and density are solved in each volume, \eqref{eq:Pressure_jump} is calculated and the interfaces are adjusted to satisfy force balance, following the original SPEC code.

\subsection{Verification of SPEC solutions with flow in single volume}
\begin{figure}[!htbp]
    \centering
    \includegraphics[width=8.5cm]{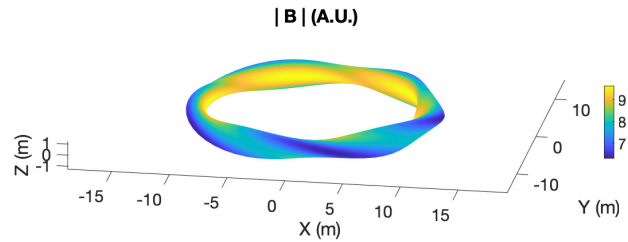}
    \caption{Plasma boundary and field strength for Case A.}
    \label{fig:caseA}
\end{figure}

In this section, we will verify the convergence of the SPEC solutions in the presence of flow.
We consider two cases with a single volume of plasma as follows.
\begin{enumerate}[{Case} A:]
    \item A classical $l=2$ stellarator with 5 field periods and field-aligned flow \cite{Loizu2016}.
    \Figref{fig:caseA} shows the plasma boundary and the corresponding field strength on the boundary.
     Parameters are $\tau = 0.01$ and $\lambda=0.01$.
    \item A tokamak with both field-aligned flow and rigid rotation.
    Parameters are $R_0=1.0$m, $a=0.3$m, $\mu=0.1$, $\tau=0.01$, $\lambda=0.1$ and $\Omega=0.1$. The boundary is circular.
\end{enumerate}
Here, $R_0$ is the major radius of the torus and $a$ the minor radius.
The error of the solution in direction $\alpha=s,\theta,\varphi$ is quantified as
\begin{equation}
    \epsilon_\alpha = \left| \left[\nabla \times \left( 1 - \frac{\lambda_i^2}{\rho} \right) \mathbf{B} - \mu_i \mathbf{B} - 2 \lambda_i \Omega_i \nabla{Z} \right] \cdot \nabla \alpha \right|,
\end{equation}
while $\bar{\epsilon}_\alpha$ defines the volume average of $\epsilon_\alpha$.

\begin{figure}[!htbp]
    \centering
    \includegraphics[width=7.5cm]{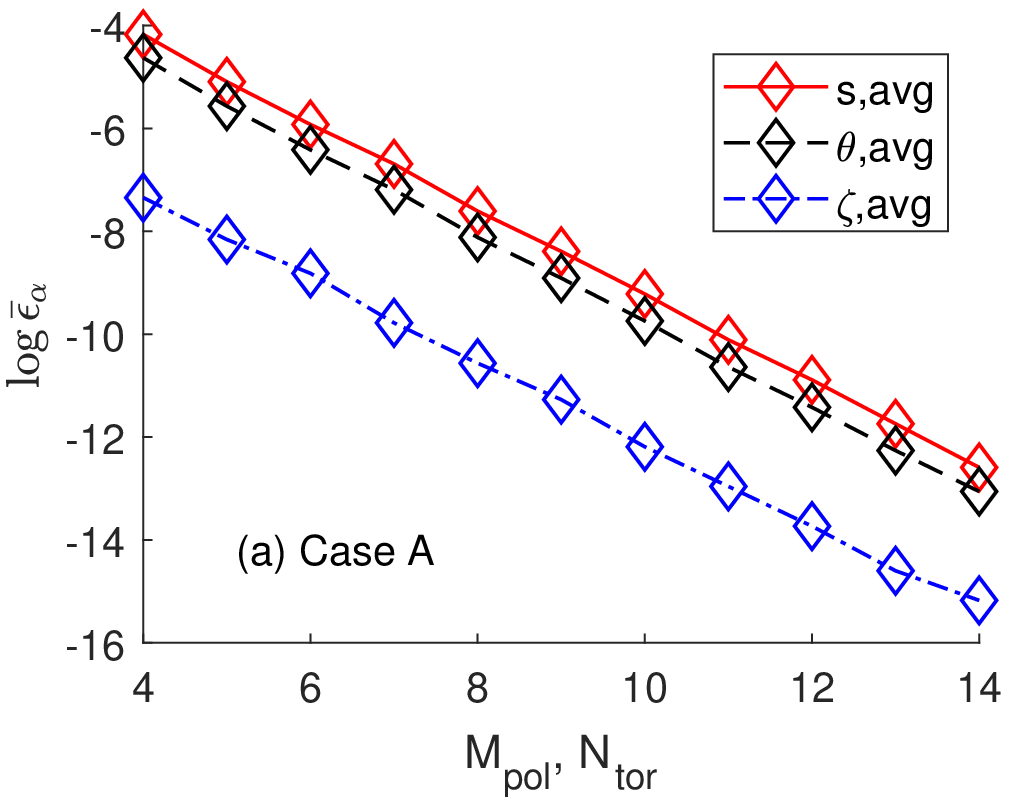}
    \includegraphics[width=7.5cm]{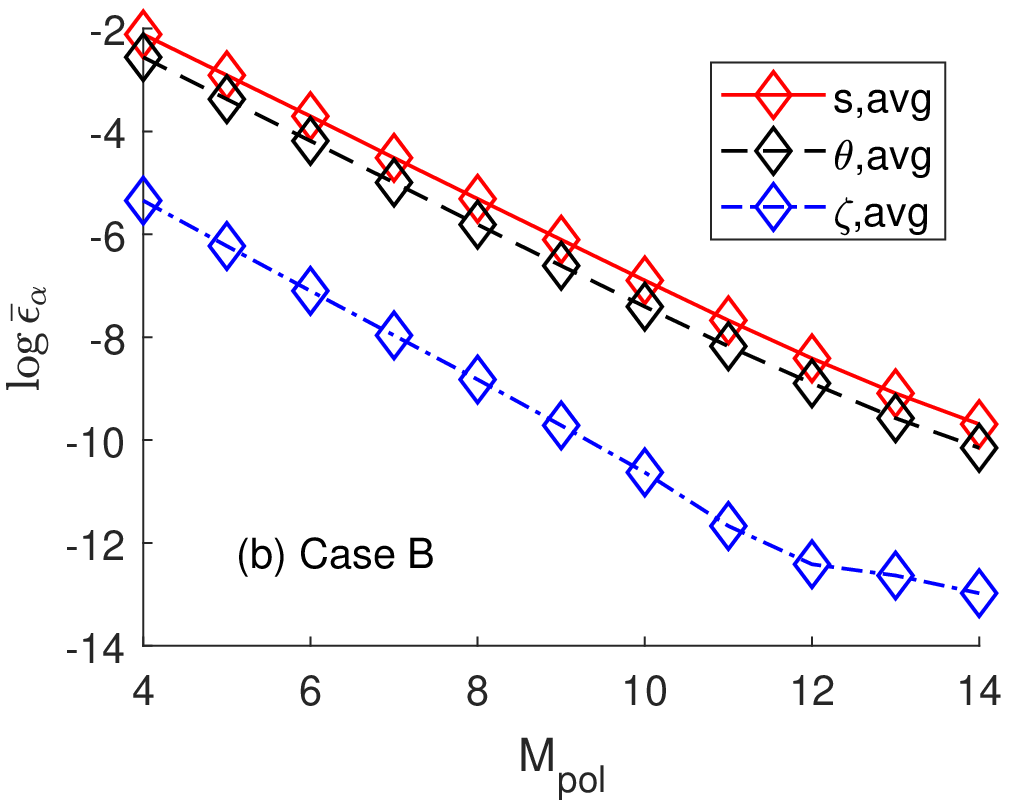}
    \caption{The force error for:
  (a) stellarator with field-aligned flow (Case A), 
  (b) tokamak with field-aligned flow and rotation (Case B).
  Radial resolution is at (a) $L_{\mathrm{rad}} = 8$, (b) $L_{\mathrm{rad}} = 10$.
  The machine precision is at $10^{-16}$.}
    \label{fig:verification}
\end{figure}

The errors of both cases are shown in \figref{fig:verification} as a function of Fourier resolution. The radial resolution is chosen to be high enough so it does not limit the precision. In case A, the error follows the trend of $\bar{\epsilon} \sim e^{-\kappa M_{\mathrm{pol}}}$ and will converge to machine precision as resolution increased. In case B, the error follows a similar trend for $M_{\mathrm{pol}} \le 12$. 
For $M_{\mathrm{pol}} > 12$, the convergence continues but at a lower rate.
The slower convergence is a consequence of the machine precision limit on the logarithmic operation in solution of the Bernoulli equation.
We found that if the Bernoulli equation is solved analytically, for instance by Taylor expansion if the parameter $M_\parallel^2 / M_{\parallel 0 }^2 - 1$ is small, the convergence will continue the trend of $M_{\mathrm{pol}} \le 12$.
To sum up, \figref{fig:verification} verifies the numerical scheme of the Beltrami solver. 

\subsection{Benchmark with a tokamak equilibrium code}
To benchmark SPEC in multiple volume setting, we compare the solution to a tokamak Grad-Shafranov equilibrium code.
The theory behind is that the continuous nested flux surfaces limit of MRxMHD should match the solution of an ideal MHD code in 2D following \secref{sec:nested_flux_surfaces}.
Existing tokamak equilibrium codes either have limited support of isothermal flux surfaces, such as FLOW \cite{Guazzotto2004}, FINESSE \cite{Belien2002} and M3D \cite{Schmitt2011},
or have no field-aligned flow, such as HELENA+ATF \cite{Qu2014}.
To benchmark with a SPEC solution, we need non-trivial modifications to these codes.
HELENA+ATF solves the modified Grad-Shafranov equation with toroidal flow and pressure anisotropy.
We added field-aligned flow into HELENA+ATF based on the Grad-Shafranov and Bernoulli equations in \appref{sec:GSE} and will not use the part related to pressure anisotropy.

We first generate a tokamak equilibrium from HELENA+ATF. The parameters used are $R_0=1$m, $a=0.3$m, on axis $\beta=0.75\%$, $F(\bar{\psi}_p) \sim 1$, $\tau(\bar{\psi}_p) \sim (1-\bar{\psi}_p)^2$, $\Omega^2(\bar{\psi}_p) \sim (1-\bar{\psi}_p)^2$,
$\lambda(\bar{\psi}_p) = 0.03(1-\bar{\psi}_p)$, $\rho_0(\bar{\psi}_p) = 1$, where $\bar{\psi}_p$ is the normalized poloidal flux given by $\bar{\psi}_p \equiv (\psi_p - \psi_{p,\mathrm{edge}})/ (\psi_{p,\mathrm{core}} - \psi_{p,\mathrm{edge}}) $.
The rotation mach number $\Omega R_0/\sqrt{\tau}$ on axis is chosen to be unity.
The SPEC interfaces are placed at flux surfaces equidistantly in $\sqrt{\bar{\psi}_p}$. The toroidal flux within each volume and the rotational transform on the interfaces are obtained from the HELENA+ATF solution and used as SPEC constraints.
The ``temperature'' $\tau_i$ within each volume is calculated as
\begin{equation}
   \tau_i \int_{\psi_{t,i-1}}^{\psi_{t,i}} d\psi_t  = \int_{\psi_{t,i-1}}^{\psi_{t,i}} \tau d\psi_t,
\end{equation}
where $\psi_t$ is the toroidal flux, 
while $\psi_{t,i-1}$ and $\psi_{t,i}$ are the enclosed toroidal flux of the bounding interfaces.
We apply similar calculations to obtain $\Omega_i$, $\lambda_i$ and $\rho_{0i}$.

\begin{figure}[!htbp]
    \centering
    \includegraphics[width=8.5cm]{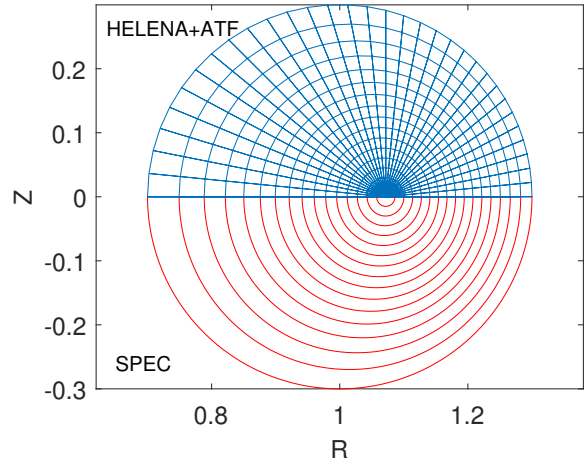}
    \caption{Flux surfaces of the Grad-Shafranov solution (HELENA+ATF) in the upper half of the figure and the SPEC interfaces with $L_{\mathrm{rad}}=4$, $M_{\mathrm{pol}}=7$ and $N_V = 16$ in the lower half. 
    Both field-aligned flow and rigid rotation are presented.}
    \label{fig:flux_surface_benchmark}
\end{figure}

\begin{figure}[!htbp]
    \centering
    \includegraphics[width=8.5cm]{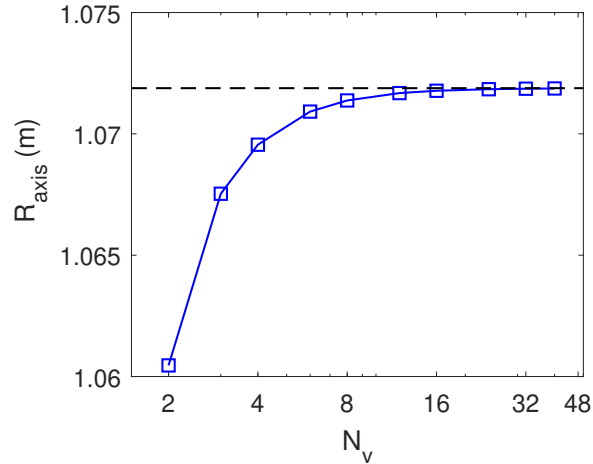}
    \caption{The position of the magnetic axis $R_{\mathrm{axis}}$ (blue square) in the innermost volume as a function of the number of volumes $N_V$. The horizontal line indicates the magnetic axis position calculated by HELENA+ATF.}
    \label{fig:raxis_benchmark}
\end{figure}

Once we set up the parameters in each volume, the position of the interfaces in SPEC are allowed to move according to the force balance
and finally rest when the energy functional reaches its stationary point.
The comparison between the HELENA+ATF solution and the final SPEC interfaces is plotted in \figref{fig:flux_surface_benchmark}, showing very good agreement.
To quantify this agreement, we start from a SPEC equilibrium with two volumes, gradually increase the number of interfaces, and record the position of the SPEC magnetic axis.
The corresponding relationship between the magnetic axis position $R_{\mathrm{axis}}$ and the number of interfaces $N_V$ is shown in \figref{fig:raxis_benchmark}.
As the number of interfaces increases, $R_{\mathrm{axis}}$ from SPEC converges to $R_{\mathrm{axis}}$ from HELENA+ATF, i.e. the ideal MHD solution.
This serves both as a benchmark of SPEC force balance and as a confirmation that MRxMHD with flow converges to ideal MHD in the limit of infinite interfaces as discussed in \secref{sec:nested_flux_surfaces}.

\begin{figure}[!htbp]
    \centering
    \includegraphics[width=8.5cm]{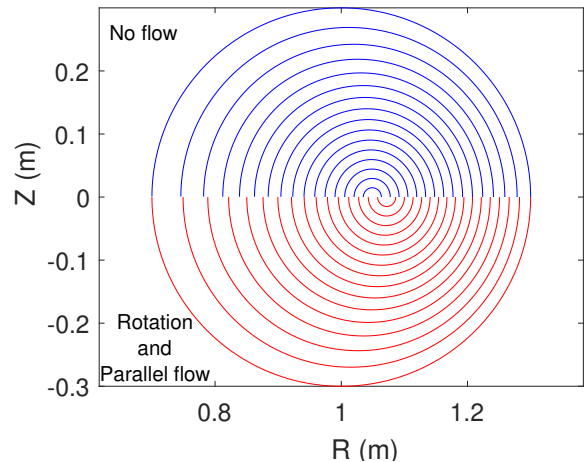}
    \caption{Interface position of the SPEC solution with both rigid rotation and field-aligned flow (lower half) and the same case with no flow. In both cases $L_{\mathrm{rad}}=4$, $M_{\mathrm{pol}}=7$ and $N_V = 16$.}
    \label{fig:flux_surface_compare}
\end{figure}

To give an idea of how flow affects the force balance, we also calculate the same case as \figref{fig:flux_surface_benchmark} but removing flow.
The comparison of the interfaces with/without flow is shown in \figref{fig:flux_surface_compare}.
It is evident from the figure that the presence of flow shifted the interfaces (flux surfaces) outward, a well studied phenomenon in the literature of tokamak equilibrium \cite{Green1972, Maschke1980, Guazzotto2004}.
Inspection of the force balance equation \eqref{eq:Bernoulli} and the Bernoulli equation \eqref{eq:Pressure_jump} shows that the density is no longer a constant within the volume and is modified by flow effects.
It is the non-homogeneity of the density on interfaces that modifies the force balance and causes the outward shift of the interfaces compared to the static case.

\section{Application to RFP plasma relaxation with flow}
\modi
\subsection{Flow profile relaxation during sawtooth crash} \label{sec:relax}
\begin{figure}[htbp]
    \centering
    \includegraphics[width=15cm]{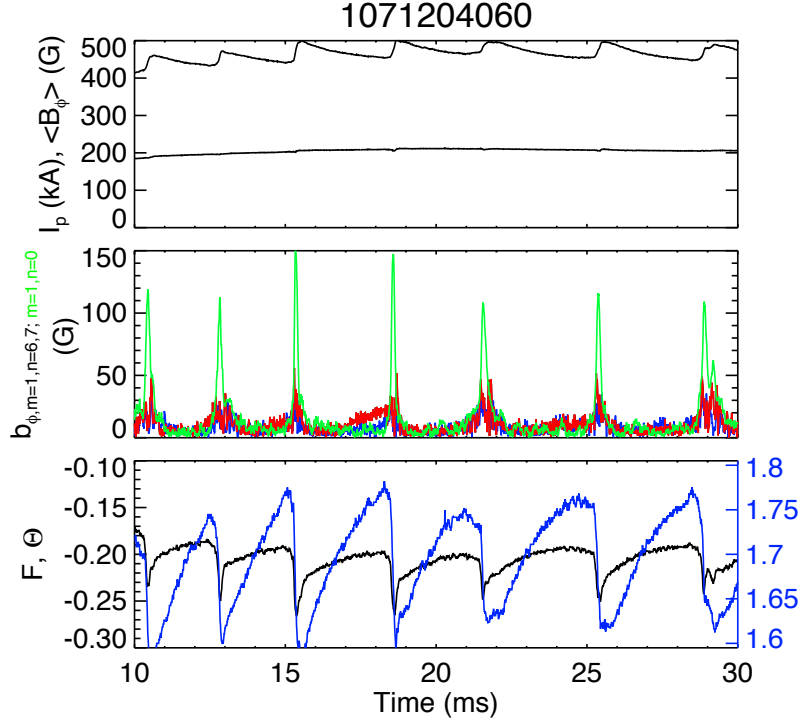}
    \caption{Time trace of MST discharge 1071204060. 
    Top panel: volume averaged toroidal field strength (upper), total current (lower).
    Middle panel: field strength of $m=1,n=6$ (red), $m=1,n=7$ (blue) and $m=0,n=1$ (green) tearing modes.
    Bottom panel: RFP parameters $F$ (black) and $\Theta$ (blue).}
    \label{fig:mst_experiment}
\end{figure}

\modii
The Madison Symmetric Torus (MST) \cite{Prager1990} is a reversed-field pinch device with major radius $R_0=1.5$m, minor radius $a=0.51$m and a circular cross section.
Sawtooth oscillations, which are associated with the reconnection events \cite{Choi2006}, are usually seen in the time series of plasma parameters (such as the total toroidal flux) in MST experiments.
These oscillations consist of a slow ramp up stage and a rapid crashing stage.
\Figref{fig:mst_experiment} shows the time traces of a few physical quantities of MST discharge 1071204060.
Take the sawtooth cycle from $15$ms to $19$ms as an example.
During the ramp up stage, the amplitude of $m=1,n=6,7$ core modes gradually grew.
At $t=18.5$ms, a global relaxation took place, resulting in the burst of $m=0$ mode.
After that, the plasma recovered to a stable state and entered the next sawtooth cycle.
The sawtooth cycle in RFP is phenomenally similar to that of a tokamak, but the detail physics is different.
In a tokamak the sawtooth is caused by a single mode due to its monotonically increasing $q$ profile.
In RFP, since the $q$ profile is monotonically decreasing and hits multiple $m = 1$ resonances with different $n$’s and also the $m=0$ resonance, multiple current-driven kink/tearing instabilities could contribute to the sawtooth oscillations \cite{Hegna1996,Nordlund1994}.
\norm

We define the \modii normalized \norm parallel flow parameter $\bar{\lambda}$ as
\begin{equation}
    \bar{\lambda} = \rho \frac{\mathbf{u} \cdot \mathbf{B}}{B^2} = \lambda + \rho \Omega R^2 \frac{\mathbf{B} \cdot \nabla{\varphi}}{B^2},
    \label{eq:bar_lambda}
\end{equation}
including the contribution from both the field-aligned flow (first term on the right hand side) and rigid rotation (second term).
Note that $\lambda$ is the field-aligned flow strength and $\Omega$ is the toroidal angular frequency according to \eqref{eq:u}.
This parameter $\bar{\lambda}$ at different stages of the sawtooth cycle was measured by Kuritsyn \etal \cite{Kuritsyn2009} and shown in Figure 4 of that reference.
The data was averaged among a class of discharges with $F=-0.2$, $\Theta=1.7$ and a total current of 200kA, with \figref{fig:mst_experiment} being a typical one of them.
Here $F\equiv B_z(a)/\langle B_z \rangle $ and $\Theta\equiv B_\theta(a)/\langle B_z \rangle$ are the RFP parameters.
The measurement showed that during the ramp up stage, $\bar{\lambda}$ had a strong peak on axis,
decreased as a function of radius, and flipped the sign half way to the edge.
During the crash, $\bar{\lambda}$ was flatten in the plasma core but changed its sign sharply at the edge.
The flattening was found in Khalzov \etal \cite{Khalzov2012} to be consistent with flow relaxation.
In this section, we intend to study this flow relaxation process using MRxMHD and SPEC with flow.
Our model differs from Khalzov \etal in three aspects:
we consider a toroidal plasma incompletely relaxed with both the cross-helicity and the angular momentum constraints, 
while Khalzov \etal considered a cylindrical plasma completely relaxed with only the cross-helicity constraint.
%we will take a multi-volume SPEC equilibrium before the sawtooth crash, compute its global helicity, cross-helicity and angular momentum,
%take them as constraints, remove all the internal interfaces to allow relaxation, and finally compute the magnetic field and flow profile after relaxation with those constraints.

\begin{figure}[htbp]
    \centering
    \includegraphics[width=7.5cm]{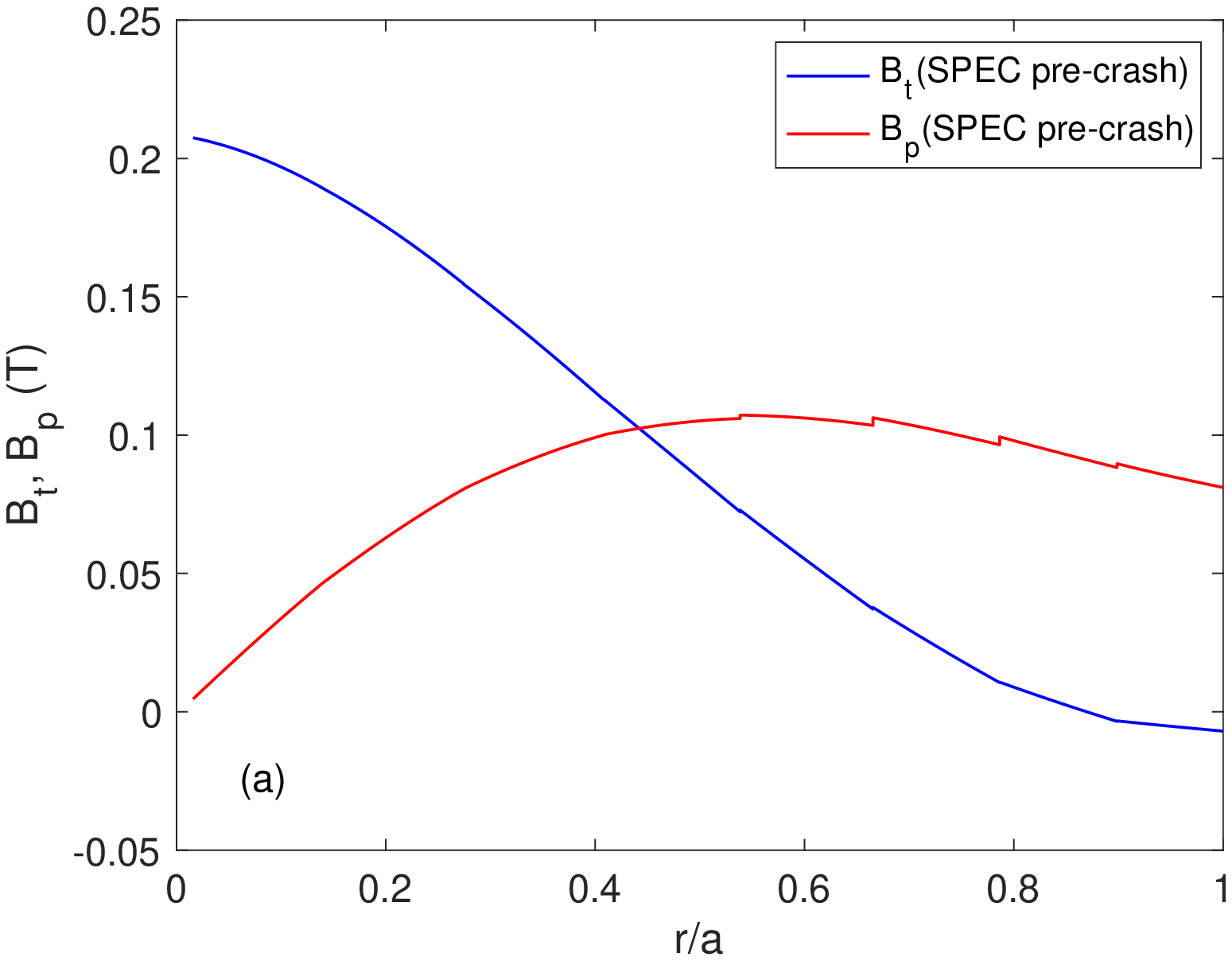}
    \includegraphics[width=7.5cm]{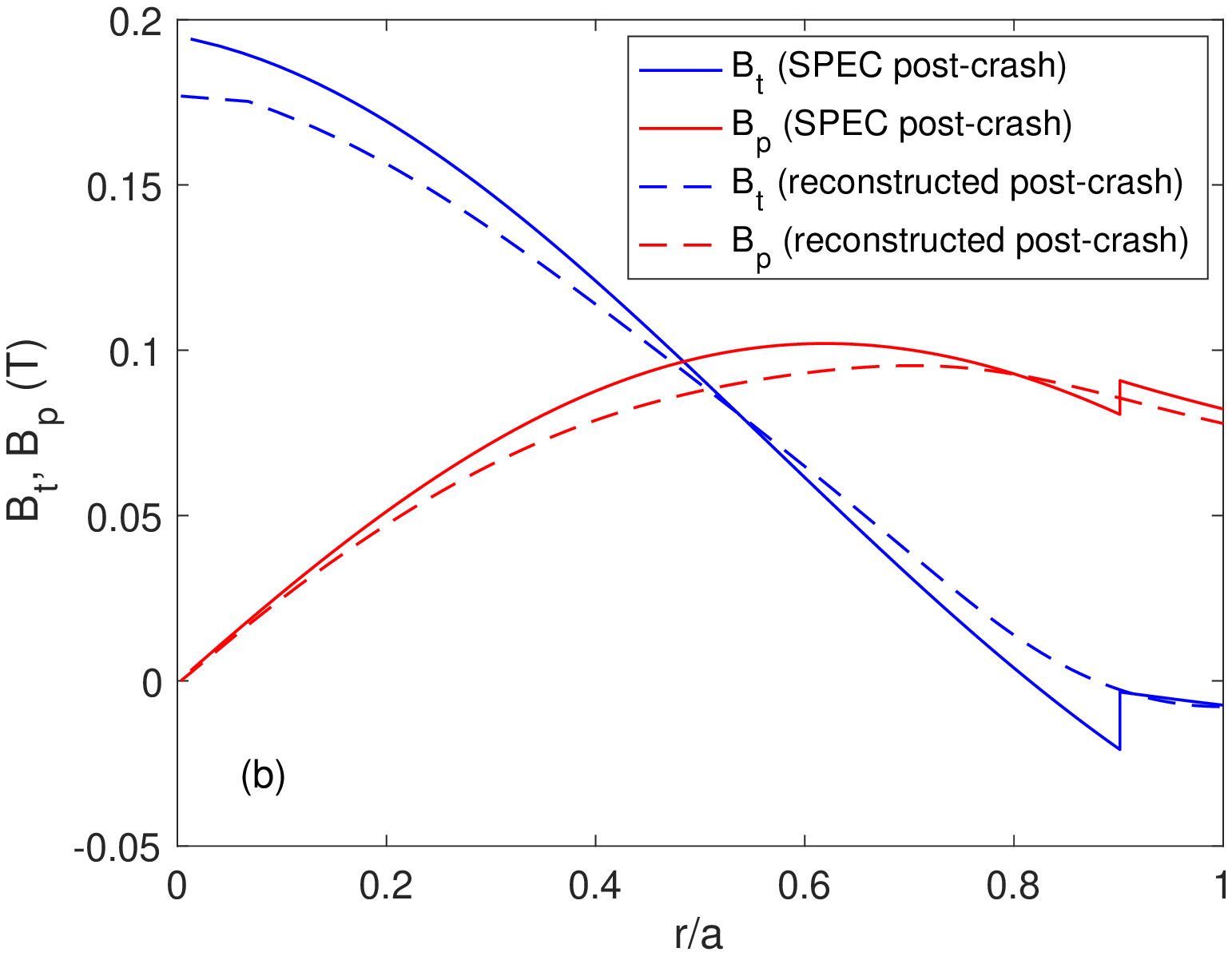}
    \caption{Toroidal field $B_t$ and poloidal field $B_p$ as a function of normalized minor radius $r/a$
     on the outbound mid-plane before and after the sawtooth crash centered at $t=18.5$ms.
    (a) SPEC pre-crash equilibrium (b) SPEC post-crash equilibrium by removing the first six interfaces and comparison to MSTFit reconstruction.
    The discontinuities in the profile suggest the placement of interfaces.}
    \label{fig:field_profile}
\end{figure}

\begin{figure}[htbp]
    \centering
    \includegraphics[width=8.5cm]{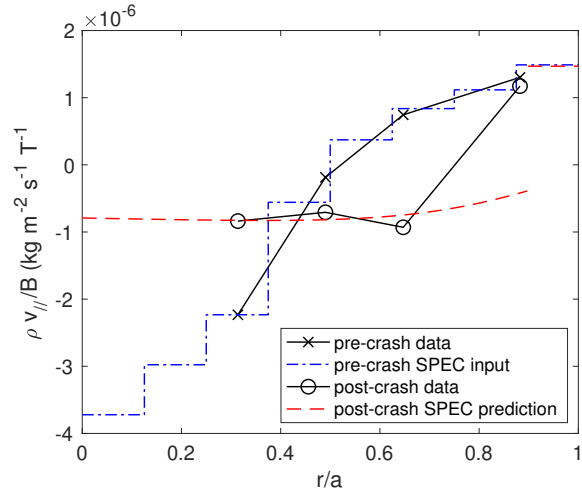}
    \caption{Comparison of the $\bar{\lambda}$ \modii (normalized parallel momentum) \norm profile between data from Kuritsyn \etal \cite{Kuritsyn2009} and SPEC input/prediction.}
    \label{fig:flow_profile}
\end{figure}

A MSTFit \cite{Anderson2004} equilibrium reconstruction of the discharge in \figref{fig:mst_experiment} was performed for a time slice just before the sawtooth crash at $t=18.5$ms.
We construct a SPEC equilibrium calculation with 8 volumes equidistantly placed in $\sqrt{\psi_p}$,
constrained volume-wisely by the magnetic helicity and fluxes computed from the original equilibrium.
The static pressure in the $i$'th volume is taken to be
\begin{equation}
    p_i = \frac{1}{\psi_{p,i+1}- \psi_{p,i}}\int^{\psi_{p,i+1}}_{\psi_{p,i}} p(\psi_p) d\psi_p,
\end{equation}
where $\psi_{p,i}$ is the poloidal flux labelling the inner interface of the $i$'th volume and $\psi_{p,i+1}$ labeling the outer surface.
For simplicity, we choose $\rho_0$ to be the experimentally measured line-averaged density 
$\rho = m_i n=1.66 \times 10^{-27} \mathrm{kg} \times 10^{13} {\mathrm{cm}}^{-3}$ and the same in all sub-volumes.
The ``temperature'' $\tau_i$ is chosen such that $p_i = \rho_i \tau_i$.
In the next step, we set the $\bar{\lambda}$ profile in each sub-volume to match piecewisely the experimental data in Kuritsyn \etal, as shown in \figref{fig:flow_profile}.
Note that Kuritsyn \etal took the tearing mode rotation velocity as a proxy for toroidal flow velocity.
As there was no resonant mode in the centre of the plasma ($r/a < 0.3$), there was no measurement of the parallel flow there.
The $\lambda$ profile at the centre is therefore extrapolated from the measurement for $r/a > 0.3$.
However, as the global constraints are volume integrals and the volume of the inner volumes are small,
our result is not sensitive to the value of $\lambda$ profile in the innermost two volumes.
The rigid rotation parameter $\Omega$ is set to zero for the pre-crash equilibrium.
\Figref{fig:field_profile}(a) gives the magnetic field components as a function of minor radius on the low field side of the mid-plane \modii before the sawtooth crash.\norm

Keeping the global helicity, cross-helicity and angular momentum as constraints, 
we remove all interfaces except the last one and recompute our equilibrium,
i.e. a relaxation is considered in the first seven volumes while the last edge volume is not relaxed.
Note that the field reversal was in the seventh volume and the interface between the seventh and the eighth volume is kept.
The resulting equilibrium magnetic fields are plotted in \figref{fig:field_profile}(b).
For comparison, the field from the MSTFit equilibrium reconstruction just after the crash is also added to the figure.
Inspection of \figref{fig:field_profile}(b) shows that the magnetic field on axis is lowered after the relaxation, agreeing with the reconstruction result qualitatively, 
but underestimating the magnitude of reduction.
The relaxation result also gives a smaller field reversal radius compared to post-crash reconstruction.
This is because the approximation we use for relaxation, which either removes an interface completely, 
or keeps it as a transport barrier and preserves the constraints before/after crash on both sides, is a too strong assumption.
In experiments, partial relaxation happens, allowing part of the helicity to leak across interfaces.
However, we will not pursue a complete match but rather keep it as a first-order coarse-grained solution.
Next, The parallel flow parameter $\bar{\lambda}$ are shown in \figref{fig:flow_profile}.
According to \figref{fig:flow_profile}, the $\bar{\lambda}$ profile has a flat top at plasma centre, 
matching the experimental observations in Kuritsyn \etal in both trend and amplitude.
The edge plasma was not relaxed and therefore, the value of $\bar{\lambda}$ stays the same as the pre-crash equilibrium.
The consistency between our relaxed flow and data implies that cross-helicity and toroidal momentum are nearly conserved quantities.

Three free parameters remain in the result we presented. 
First, the $\lambda$ profile of the pre-crash equilibrium in the plasma core is extrapolated and therefore bares uncertainties.
Second, the density profile is taken to be constant in our calculation, while in experiment it is flat in the plasma core and decreases at the plasma edge.
Finally, the unrelaxed interface at the edge is placed at approximately $7/8$ of the plasma minor radius, a seemly arbitrary choice.
We investigated all three parameters, noting that changing the last one has the most significant impact on the final flow amplitude.
For example, if the position of the unrelaxed interface $r_{\mathrm{ui}}$ moves to $r_{\mathrm{ui}}=a$, i.e. all regions are relaxed, the final flow amplitude will reduce to about half of the measured value.
Nevertheless, the allowed parameter range of $r_{\mathrm{ui}}$ is determined considering that
\begin{enumerate}
    \item the relaxed region should contain the field reversal of the pre-crash equilibrium at $r/a=0.87$, as indicated by the large $m=0$ mode amplitude during the reconnection event; 
    \item and that the most outward data point at $r/a=0.9$ should be in the unrelaxed region, as the measured $\bar{\lambda}$ is nearly unchanged before/after the crash.
\end{enumerate}
Consequently, $r_{\mathrm{ui}}/a$ falls in to a small range of $0.87<r_{\mathrm{ui}}/a<0.9$, 
within which the final value of $\bar{\lambda}$ after relaxation does not change significantly.
The result presented in \figref{fig:flow_profile} has $r_{\mathrm{ui}} \approx 0.88$ and is therefore justified.
\norm

\subsection{3D helical RFP equilibrium with flow} \label{sec:3D}
\modi
The magnetic fields of RFP plasmas have very rich 3D structures and these structures can be reproduced by SPEC.
In Dennis \etal \cite{Dennis2013}, a minimally constraint model was built by slicing a RFP VMEC \cite{Hirshman1986} equilibrium into two sub-volumes.
The fluxes and magnetic helicity in each volume is then computed and used as inputs to SPEC.
Two equilibria were found with the same set of constraints, one being axisymmetric, the other being helical.
It was discovered when the interface is placed near the core, 
the helical equilibria can have a lower MHD energy than the axisymmetric one and is therefore a more energy favoured state.
These steps can be repeated for a MST equilibrium with finite plasma flow.
\norm

\begin{figure}[!htbp]
    \centering
    \includegraphics[width=8.5cm]{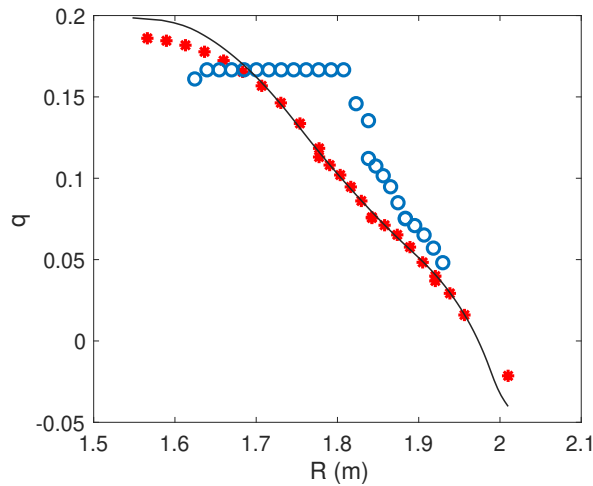}
    \caption{The safety factor $q$ as a function of outbound major radius $R$ of the static axisymmetric VMEC equilibrium (solid black line), 
    the axisymmetric SPEC equilibrium with flow (red stars), and the helical bifurcated SPEC equilibrium with flow (blue circles). 
    The SPEC safety factor is calculated by field line tracing.}
    \label{fig:qprofile}
\end{figure}

We reproduce a MSTFit equilibrium reconstruction of the discharge in \figref{fig:mst_experiment} at a time slice 1ms before sawtooth crash using VMEC  \cite{Hirshman1986}.
The $q$ profile of this equilibrium is plotted in \figref{fig:qprofile}.
We will not use the same pressure profile as the reconstruction since it has a strong off-axis peak, indicating possible helical structures.
Instead, we set the pressure profile to be $p=p_0(1-\psi_p^2)$, with the pressure on axis $p_0=0.5$kPa consistent with the reconstruction.
Following Dennis \etal, we slice the plasma volume into four sub-volumes at certain VMEC flux surfaces.
These flux surfaces should have a irrational safety factor.
The first flux surface is chosen to be the $q_{8/9}=(1+\gamma_g)/(8+9\gamma_g)$ flux surface, 
in which $\gamma_g = (1+\sqrt{5})/2$ is the golden ratio and $q_{8/9}$ is the ``most noble''  \cite{Hudson2004} irrational number between $1/8$ and $1/9$.
In other words, the continuous fraction expansion of $q_{8/9}$ has the most number of one's and its flux surface has the largest possibility to survive under non-axisymmetric perturbations.
The second and third flux surfaces are selected in exactly the same way but for $q_{12/13}$ and $q_{24/25}$.
Similar to Dennis \etal  \cite{Dennis2013}, the number of interfaces and their safety factor will have a strong effect on the final equilibrium.
In particular, the existence of an interface creates a shielding current which heals islands and chaos and promotes good flux surfaces around it.
For a physics study, one will need to take this into account carefully, 
such as constructing a sequence of equilibria for the sawtooth cycle and allowing the interfaces to break one by one as the island grows.
However, in this work we will not pursue this path,
but rather construct an equilibrium with less rigorous choice of interfaces as described above.
The purpose is to demonstrate the capacity of the tool on which
future dedicated research can be built.

After sub-volumes are chosen, the toroidal flux, poloidal flux and helicity in each volume are calculated from the VMEC equilibrium.
The field period is set to NFP=6 to stay consistent with the experimental observation that the dominant mode had the mode number $m=1$, $n=6$.
These constraints are used to drive an SPEC equilibrium calculations with the normalization from SI unit to SPEC given by $B/\sqrt{\mu_0} \rightarrow B$,
where $\mu_0$ is the vacuum permeability.
Parallel flow is now added to the equilibrium, with the $\lambda$ profile resembles the experimental measurement.
The discretized $\lambda/m_i$ value is taken to be $-75,0,40,60$ in the unit of $10^{22} \mathrm{m}^{-2} \mathrm{s}^{-1} \mathrm{T}^{-1}$ from the most inner volume to the most outer volume,
to match the measured profile in Figure 4(b) of Kuritsyn \etal. 

\begin{figure}[!htbp]
    \centering
    \includegraphics[width=7.5cm]{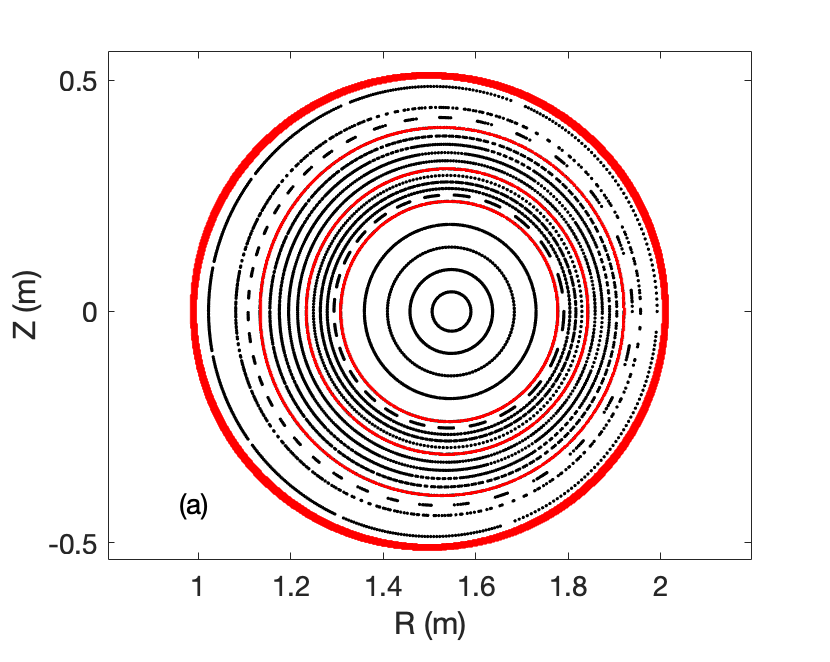}
    \includegraphics[width=7.5cm]{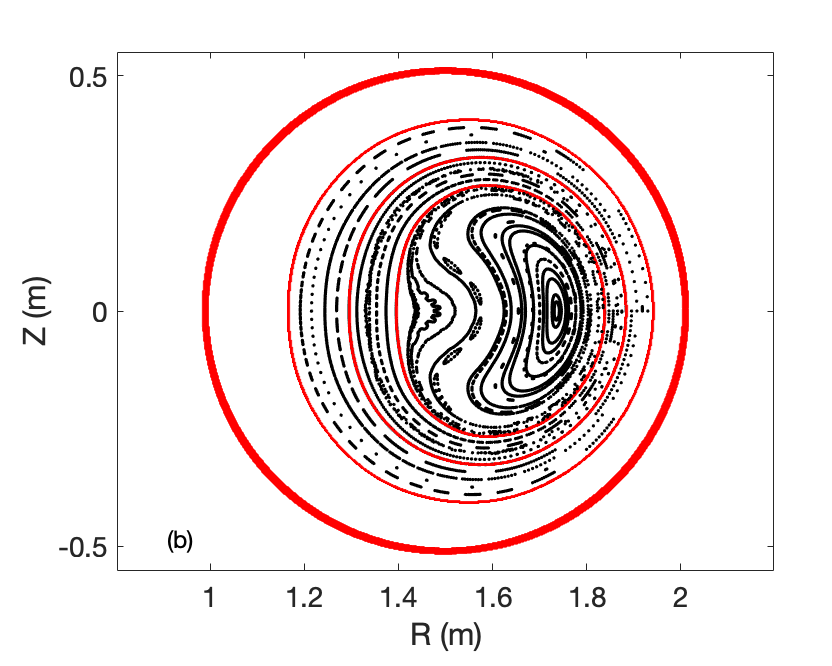}
    \caption{The \Poincare cross section at $\varphi=0$ for (a)axisymmetric solution and (b)helical solution.
    The red solid lines indicates the position of the interfaces. 
    No \Poincare plot was shown in the outermost sub-volume of the helical case because the very small toroidal field in that region.}
    \label{fig:poincare}
\end{figure}

One equilibrium solution satisfies the aforementioned constraints has axisymmetric magnetic field with nested flux surfaces.
The $q$ profile of this equilibrium is plotted in \figref{fig:qprofile} and is shown to match the VMEC solution.
The \Poincare section of this equilibrium is plotted in \figref{fig:poincare}(a).
Keeping the same constraints, SPEC found another equilibrium with a $m/n=1/6$ helical core that has lower total energy,
with \figref{fig:poincare}(b) being its \Poincare section.
The $q$ profile of the helical equilibrium is over-plotted in \figref{fig:qprofile},
showing a plateau of $q=1/6$ in the plasma core due to the helical core structure.
\modi
The safety factor $q$ here is calculated with respect to the centre of the innermost region.
The construction of another safety factor by choosing the axis to be the centre of the helical core is also possible but will not be pursued here.
\norm
The existence of a lower energy state indicates that the axisymmetric equilibrium
corresponds either to a saddle point or a local minimum of the energy functional.
It is known that the RFP configurations are vulnerable to current-driven kink/tearing instabilities which forms the sawtooth cycle  \cite{Watt1983},
and the lower energy state is possibly an equilibrium with saturated instabilities.

\begin{figure}[!htbp]
    \centering
    \includegraphics[width=8.5cm]{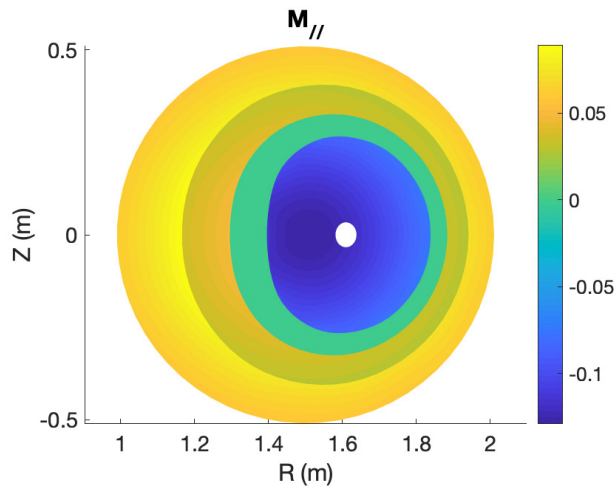}
    \caption{The contour of the parallel Mach number $M_\parallel$ for the helical equilibrium.}
    \label{fig:mparallel}
\end{figure}

The contour of the parallel Mach number $M_\parallel$ on the $\varphi=0$ surface for the helical equilibrium is shown in \figref{fig:mparallel}.
It takes four steps ranging from $-0.13$ to $0.08$ due to our approximate four volume equilibrium.
\modii The range of $M_\parallel$ in the axisymmetric state is similar but not plotted here. \norm
We note that the density scales as $\rho = \rho_0 \exp(-M_\parallel^2/2)$
and in this case, since $|M_\parallel| \le 0.13$, the change of density due to the Bernoulli equation is less than $1\%$.
Furthermore, The global plasma beta is about $4\%$ and as a consequence, the $\lambda^2/\rho$ factor in the Beltrami equation, 
which stands for the squared \Alfven Mach number, is in the order of $10^{-4}$.
Therefore, the field-aligned flow in the parameter range of the modeled experiment does not have a strong effect both on the Beltrami equation and in the force balance condition \eqref{eq:Pressure_jump}.
Nevertheless, a self-consistent numerical equilibrium for RFP with field-aligned flow in toroidal geometry is produced for the first time.
This equilibrium, although coarse-grained in four volumes, stands as a baseline for future studies with more delicate choices of interfaces as mentioned earlier in this section.

\section{Conclusion} \label{sec:conclusion}
In this paper, we revisited the Multi-region Rela\underline{x}ed MHD MHD (MRxMHD) theory
with cross-helicity and angular momentum constraints, \modii in addition to the original magnetic helicity constraint, \norm to capture flow effects.
The stationary point of our MRxMHD energy functional gives equilibria with stepped field-aligned flow and rotation,
in additional to the stepped pressure and parallel current profiles across different sub-volumes.
They are mathematically well-defined solutions in 3D that allow the co-existence of flux surfaces, islands and chaos,
while the effect of flow is self-consistently cooperated. 
We implemented these new features into the Stepped Pressure Equilibrium Code (SPEC),
which can produce numerical equilibrium solutions in 3D with flow.
We verified the convergence of the numerical scheme 
and compared the infinite interfaces limit of the SPEC-flow solution to that of a tokamak equilibrium code assuming nested flux surfaces, showing very good converged agreement.
\modi
The newly developed tool was then used to model equilibria of the MST reversed-field pinch experiment with a non-zero flow profile.
We constructed a post-sawtooth-crash SPEC equilibrium by relaxing the core volumes of a pre-crash one, 
keeping the magnetic helicity, the cross-helicity and the angular momentum constraints.
The resulting equilibrium has a parallel flow profile matching that of experimental observations. 
The flattening of the parallel flow profile found here is consistent with the theoretical cylindrical results \cite{Khalzov2012}.
However, this is the first time that the MST experimental equilibria is coupled to a 3D toroidal equilibrium code for the
prediction of the relaxed states under specific constraints.
\norm
Finally, we gave an example of a experimentally relevant equilibrium of a MST discharge: 
the equilibrium contains a 3D helical core and has a lower MHD energy than its axisymmetric counterpart.

\modii 
We plan to pursue these promising preliminary results, and further extend these analysis to compare with the experiments.
\norm
The most straightforward extension of the current paper will be to make use of our new version of SPEC, 
construct a sequence of flowing equilibria that tracks the magnetic field structure of a RFP sawtooth cycle,
and compare to tomographic results \cite{Franz2006}.
As mentioned in \secref{sec:relax}, one will need to choose a number of interfaces at irrational rotational transform at the beginning of the sawtooth cycle, track the saturated instabilities as current ramps up,
and remove interfaces when they break.
Next, some interest remains in examining the transition between sub/supersonic flow,
and to go beyond the sub-Alfven regime in which the Beltrami equation becomes hyperbolic.
Furthermore, our new tool solves for the (unstable) initial or steady state of the time-dependent MRxMHD theory on the basis of which future simulation will be performed.
Lastly, we hope to diversify the type of flow allowed in our relaxation theory,
by adding and/or editing the global constraints in different physical scenarios,
to capture more classes of flowing equilibrium,
such as that with an undamped flow in the symmetry direction of a quasi-symmetric stellarator.

\ack
The authors would like to thank Drs. J Loizu (EPFL), CX Zhu (PPPL) and N Sato (Kyoto University) for useful suggestions.
The authors are also grateful to National Computational Infrastructure Australia for computational resources.
This work is funded by Australian ARC projects DP140100790 and DP170102606, and the U.S. Department of Energy.
This work was supported by a grant from the Simons Foundation/SFARI (560651, AB).
The first author would like to thank National Fusion Research Institute (NFRI) for travel support to present this work at NFRI.

\appendix 
\section{Grad-Shafranov-Bernoulli Equation with poloidal and toroidal flow}
\label{sec:GSE}
In an axisymmetric toroidal plasma, the magnetic field can be written as
\begin{equation}
    \mathbf{B} = \nabla \psi_p \times \nabla \varphi + R B_\varphi \nabla \varphi,
\end{equation}
with $\psi_p(R,Z)$ the poloidal flux function and $B_\varphi$ the magnetic field strength in toroidal direction.
The form of the flow is given by
\begin{equation}
    \mathbf{u} = \frac{\lambda (\psi_p)}{\rho} \mathbf{B} + \Omega(\psi_p) R^2 \nabla \varphi
    \label{eq:appendix_u}
\end{equation}
The isothermal Bernoulli Equation is derived by taking the $\mathbf{B}$ direction of the momentum equation \eqref{eq:equation_of_motion}.
It as the form
\begin{equation}
    T(\psi_p) \ln \rho + \frac{\lambda^2 B^2}{2\rho^2} - \frac{1}{2}\Omega^2 R^2 = H(\psi_p).
    \label{eq:appendix_Bernoulli}
\end{equation}
The temperature $k_B T/m_i$ is exchangeable with the Lagrange multiplier $\tau$ in the MRxMHD theory.
The $\varphi$ direction of \eqref{eq:equation_of_motion} gives
\begin{equation}
    B_\varphi R = \frac{F(\psi_p) + \lambda \Omega R^2}{1 - \lambda^2/\rho}.
\end{equation}
And finally, the $\nabla \psi_p$ direction of \eqref{eq:equation_of_motion} leads to the modified Grad-Shafranov equation given by
\begin{eqnarray}
\fl    \nabla \cdot \frac{1}{R^2}\left(1 - \frac{\lambda^2}{\rho} \right) \nabla \psi_p = - \frac{B_\varphi}{R} F' - (\mathbf{u} \cdot \mathbf{B}) \lambda' \nonumber\\
    - \rho \left(u_\varphi R \Omega' + H' + T' - T' \ln \rho \right). 
\end{eqnarray}
Consequently, the system is determined by five flux functions $\{F(\psi_p),T(\psi_p),H(\psi_p),\Omega(\psi_p),\lambda(\psi_p)\}$ and the boundary condition (in the fixed-boundary case $\psi_p=$ constant on the plasma boundary).
\section*{References}
\bibliography{ref}
\end{document}